\documentclass[11pt,twoside]{article}
\usepackage{asp2014}

\usepackage[vpos=0.05\paperheight,angle=0.0,hpos=0.2\paperwidth,color=black,alignment=l]{draftwatermark}
\SetWatermarkText{
\textbf{\LARGE Preprint}\\
\textit{Astronomical Data Analysis Software and Systems XXX}\\
\textit{J.E. Ruiz, F. Pierfederici, and P. Teuben, eds.}\\
\textit{ASP Conference Series}\\
\textit{\copyright 2021 Astronomical Society of the Pacific}
}
\SetWatermarkScale{1}
\SetWatermarkFontSize{0.3cm}

\aspSuppressVolSlug
\resetcounters

\newcommand{\dask}{\textsc{Dask }}
\newcommand{\daskms}{\textsc{Dask-ms }}

\newcommand{\numpy}{\textsc{Numpy }}

\begin{document}

\title{Tricolour: an optimized SumThreshold flagger for MeerKAT}

\author{Benjamin~V.~Hugo$^{1, 2}$, 
        Simon~Perkins$^{1}$, Bruce~Merry$^{1}$,
        Tom~Mauch$^{1}$ and Oleg~M.~Smirnov$^{2,1}$}
\affil{$^1$South African Radio Astronomy Observatory, Cape Town, South Africa; \email{bhugo@ska.ac.za}}
\affil{$^2$Rhodes University, Makhanda (Grahamstown), Eastern Cape, South Africa}

\paperauthor{Benjamin~V.~Hugo}{bhugo@ska.ac.za}{0000-0002-2933-9134}{South Africa Radio Astronomy Observatory}{Radio Astronomy Research Group}{Cape Town}{Western Cape}{7925}{South Africa}




\begin{abstract}
We present \textsc{Tricolour}, a package for Radio Frequency Interference mitigation of wideband finely channelized MeerKAT correlation data. The MeerKAT passband is heavily affected by interference from satellite, mobile, aircraft and terrestrial transponders. Coupled with typical data rates in excess of 100 GiB/hr at 208kHz channelization resolution, mitigation poses a significant processing challenge. Our flagger is highly configurable, parallel and optimized, employing Dask and Numba technologies to implement the widely used SumThreshold and MAD interference detection algorithms. We find that typical 208kHz channelized datasets can be processed at rates in excess of 400 GiB/hr for a typical L-band flagging strategy on a modern dual-socket Intel Xeon server.
\end{abstract}

\section{Radio Frequency Interference impacts on MeerKAT observation}

\begin{figure}[h]
    \centering
    {\includegraphics[width=0.5\textwidth]{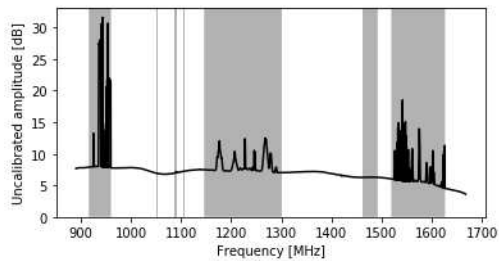}}
    {\caption{Mean visibility amplitude of a typical unflagged and uncalibrated L-band MeerKAT observation. Darkened areas show subbands where persistent interference is most likely, and  serves as a ``static'' background mask during application of mitigation routines.}
    \label{fig:MKLband}}
\end{figure}

\textit{Radio Frequency Interference (RFI)} can be defined as any interfering signal that negatively affects the observable instrumental bandwidth of a radio telescope. The most prominent RFI impacting cm wavelength observations using MeerKAT \citep{jonas2018meerkat} are spatially coherent man-made telecommunications and navigational satellite transmissions. These substantially impacts all observation, particularly that of L-band, shown in Fig.~\ref{fig:MKLband} for a typical observation.

RFI affecting large portions of the bandwidth presents a substantial risk for, especially, sensitive continuum radio science projects, because the sensitivity of such observations scale as the square root of the observable bandwidth. As shown in Fig.~\ref{fig:MKLband}, man-made interference power is orders of magnitude above the relatively dominating sub-Jy \textit{Active Galactic Nuclei} (AGN) population at cm wavelengths --- even on a bright calibrator field! A large number of science cases require very long observations to target imaging sensitivities at the level of only a few $\mu\text{Jy}/\text{beam}$ RMS noise, e.g. \citet{Mauch_2020}. 
Multi-hour observations using all available bandwidth is only one factor determining image observation sensitivity --- another important factor is the number of unique interferometer spacings (baselines) contributing to the final synthesized image. This scales, roughly, quadratically with the number of antennae in the interferometric array. 


The dump rate of MeerKAT's raw correlated visibilities (excluding metadata) is 0.14 and 1.10 TiB~/~hr for coarse and fine channelization respectively, at a typical dump rate of 8s. RFI mitigation on a large wideband sensitive array therefore poses a significant processing challenge, requiring a flexible and parallel implementation of mitigation routines.

\section{RFI detection and mitigation}
Outlier detection routines are commonly used to detect and mitigate RFI in interferometry. The results are stored in a spectro-polarimetric flag array with the same shape as the data column, which is used in further calibration and imaging of the radio product. 

Mitigation of man-made RFI is substantially simplified by two properties of the measured signal. Firstly, transmission signals are ordinarily highly polarized. Much of the bright cm dominating radio AGN population is linearly polarized to a couple of percent. Secondly, a radio interferometer measures phase information of the spatial locality of emitting sources. The complex vector rotates quickly on long spacings for all declinations away from the celestial poles. Through complex averaging, a substantial portion of the detectable RFI is washed out.

Various simple and effective outlier detection techniques that can be used to detect the remaining RFI not washed out in the fringes of the interferometer, are discussed in \citet{offringa2010post}. The \textsc{SumThreshold} algorithm is a simple and fast windowed procedure widely used for RFI detection that can be applied to two dimensions --- that of time and frequency per spacing. The method clips values where the sum of the values within a time and channel window exceeds an interatively adjusted average threshold. The method is very sensitive to RFI while maintaining good false-positive classification rates for large enough window sizes. The algorithm is trivially parallel over the unique interferometer spacings (baselines), but requires a reordering step to select the time and frequency subsets per baseline from data that is traditionally stored in time-sorted row order. For a large array such as MeerKAT this step is amortized by the dominating run time of the sum-thresholding for various window sizes and number of iterations. To aid in background estimation on short spacings, we provide options to apply ``static'' masks to flag large bands of persistent RFI (such as the one shown in Fig.~\ref{fig:MKLband}) on adjustable ranges of baseline length, as well as options to form residual products based on predicted sky models. 

\begin{figure}
    \centering
    \begin{minipage}{0.95\textwidth}
         \centering
         \includegraphics[width=0.325\textwidth]{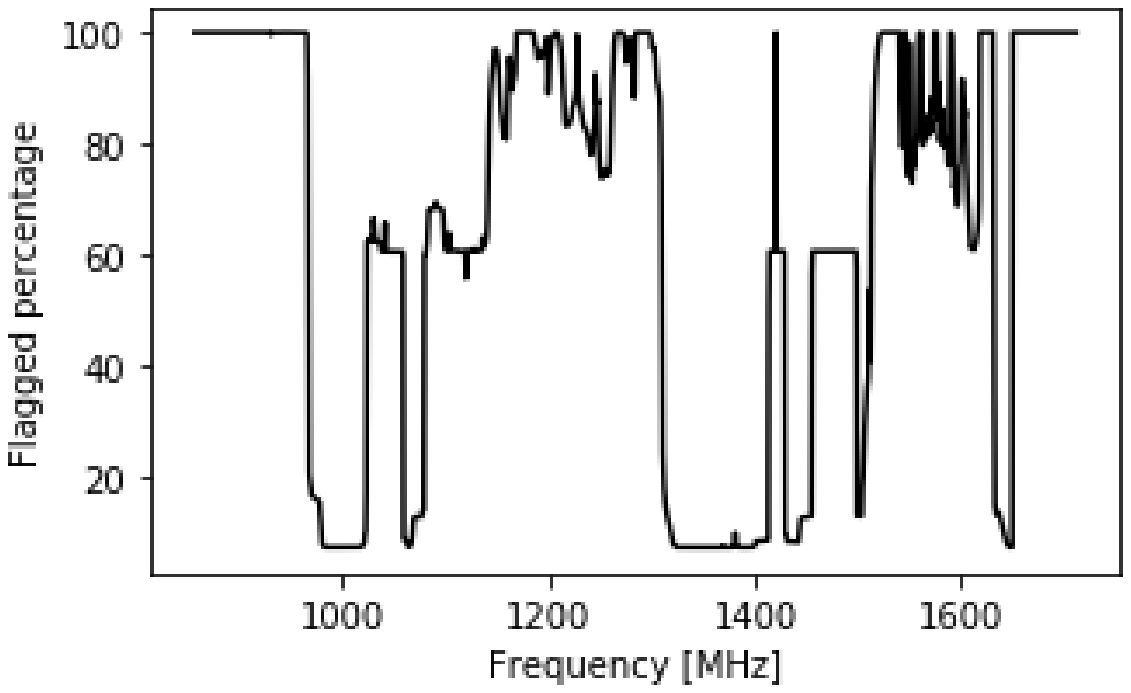}
         \includegraphics[width=0.325\textwidth]{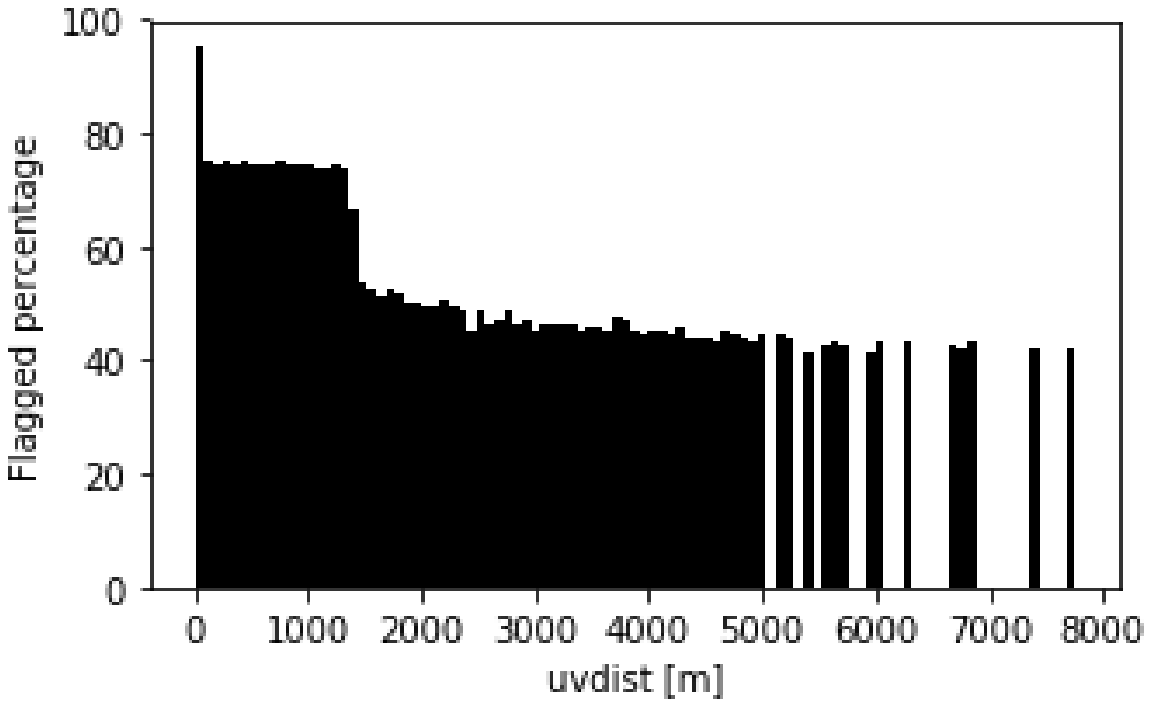}
         \includegraphics[width=0.325\textwidth]{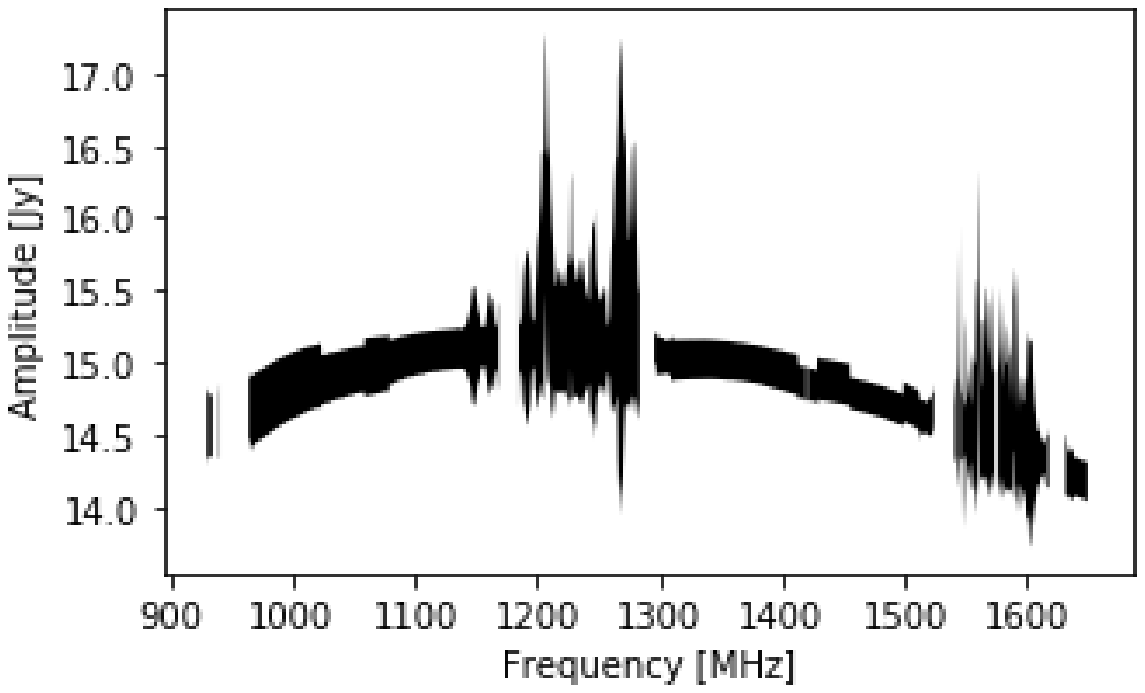}
     \end{minipage}
    \begin{minipage}{0.95\textwidth}
         \centering
         \includegraphics[width=0.325\textwidth]{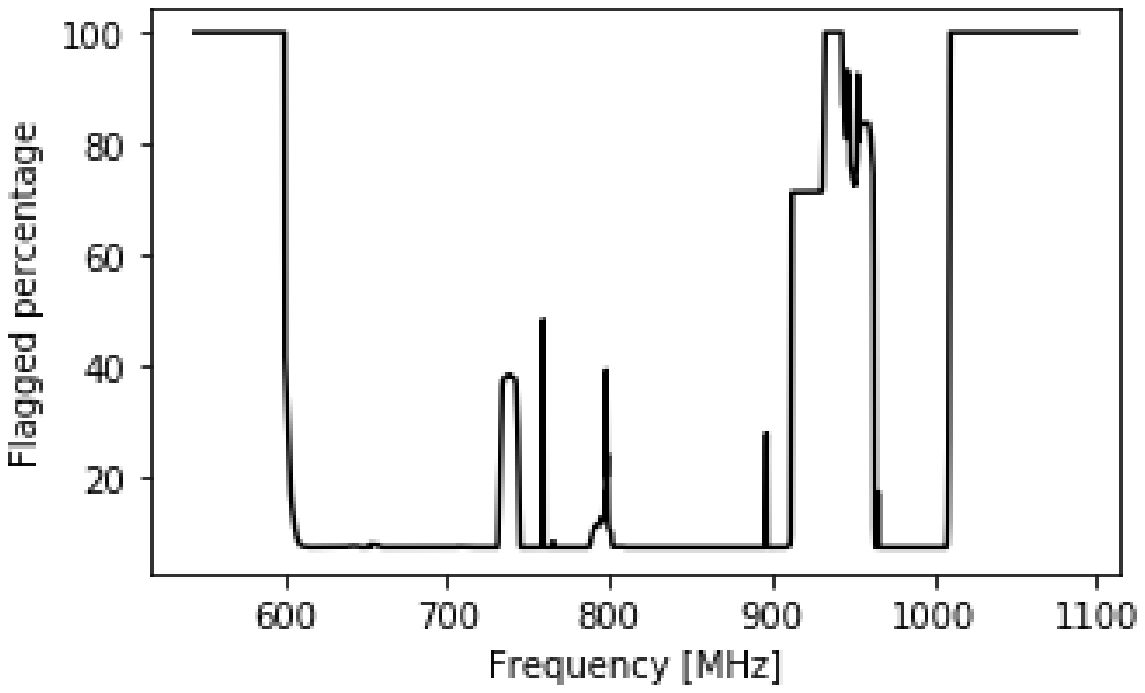}
         \includegraphics[width=0.325\textwidth]{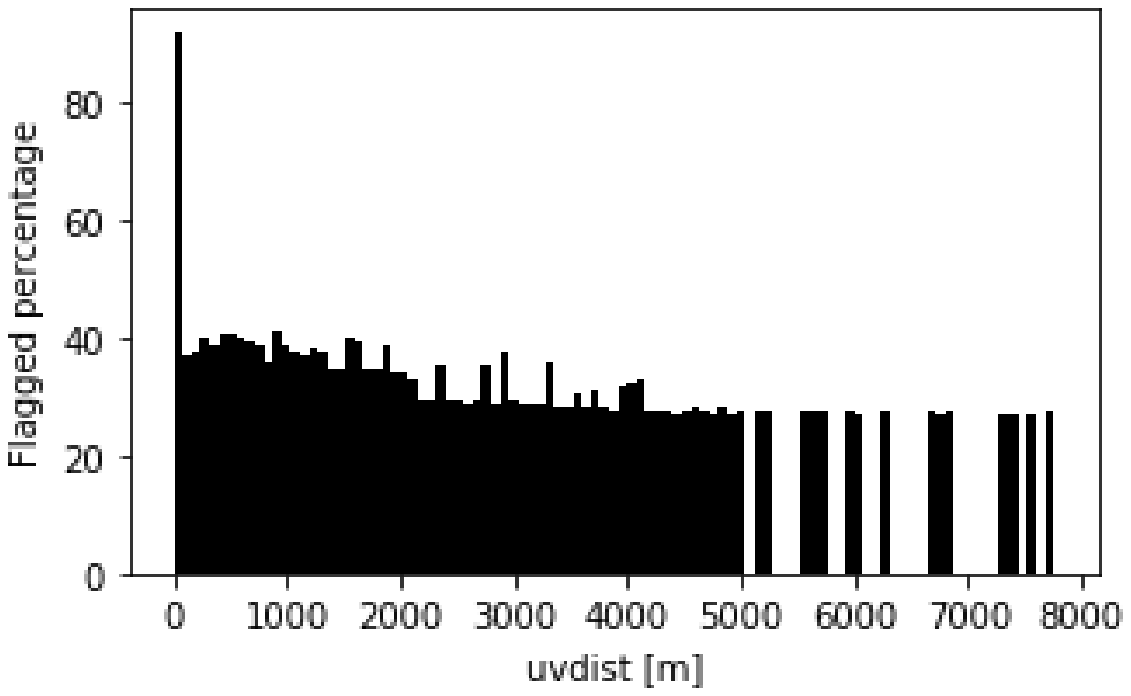}
         \includegraphics[width=0.325\textwidth]{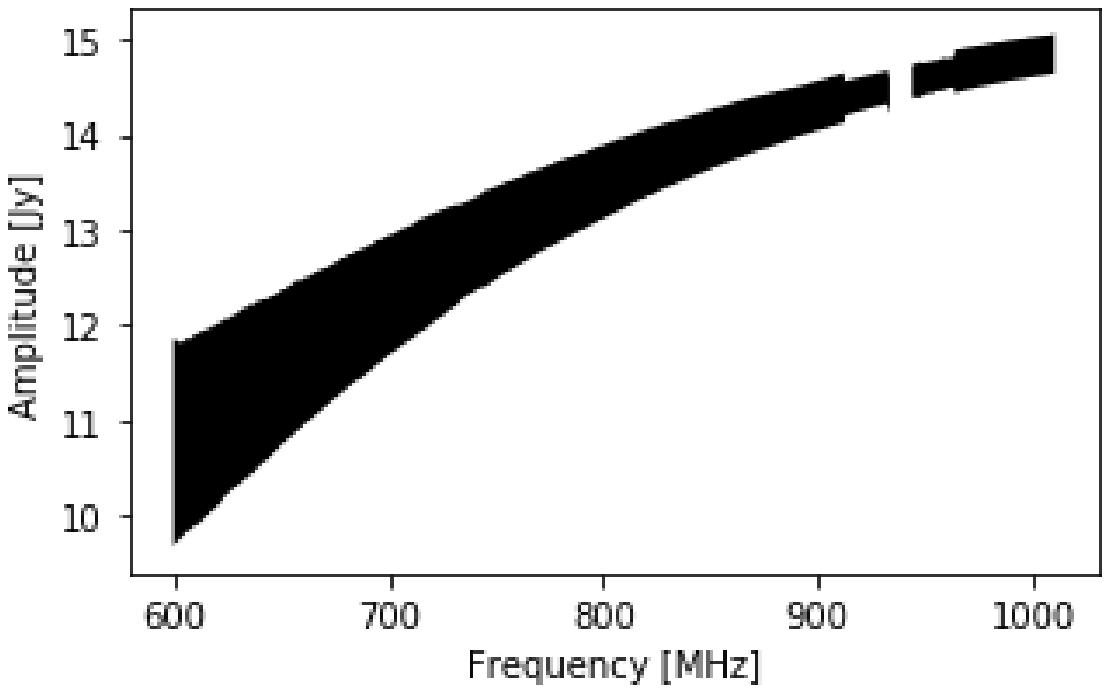}
     \end{minipage}
    \caption{Here we show bandwidth and spatial flagging statistics after applying typical multi-pass flagging strategies for L (first row) and UHF (second row) bands of calibrator PKS B1934-638. UHF band is generally very clean compared to L-band, however we note that the large spread in visibility amplitudes at the bottom of the band is due to the substantial cumulative contribution from the off-axis AGN population, due to the wide field of view of the instrument. The contribution from navigational systems is substantially reduced in comparison to what was seen in Fig.~\ref{fig:MKLband}}
    \label{fig:flag_stats}
\end{figure}

The sizes of the windows, as well as the sensitivity adjustment parameter, have to be fine tuned by hand for the particular science case and channelization regime of the observation, to minimize false positives. To achieve this we provide a \texttt{YAML}-based interface to the user to allow users to customize strategies to their requirements. Typical flagging statistics and results for MeerKAT are shown in Figure~\ref{fig:flag_stats}.

\section{A parallel, scalable implementation}
Our flagger is distributed as an Open Source \textsc{Python} package, available on the Python Package Index as ``\texttt{tricolour}". We have implemented the \textsc{SumThreshold} method written in \numpy \citep{Harris2020} and cache-optimized  \textsc{Numba} \citep{lam2015numba} kernels. The user-available version of the flagger ingests data from Measurement Set v2.0 columns exposed as chunked \dask Arrays by \daskms. This architecture is more fully described in \citet{O8-131_adassxxx}. A trimmed version of the flagger is used in the online MeerKAT Science Data Processor which flags most of the substantial RFI in buffered raw data at line rates, before storage to the MeerKAT archive.

\begin{figure}
    \centering
    \begin{minipage}{0.36\textwidth}
         \centering
         \includegraphics[width=0.99\textwidth]{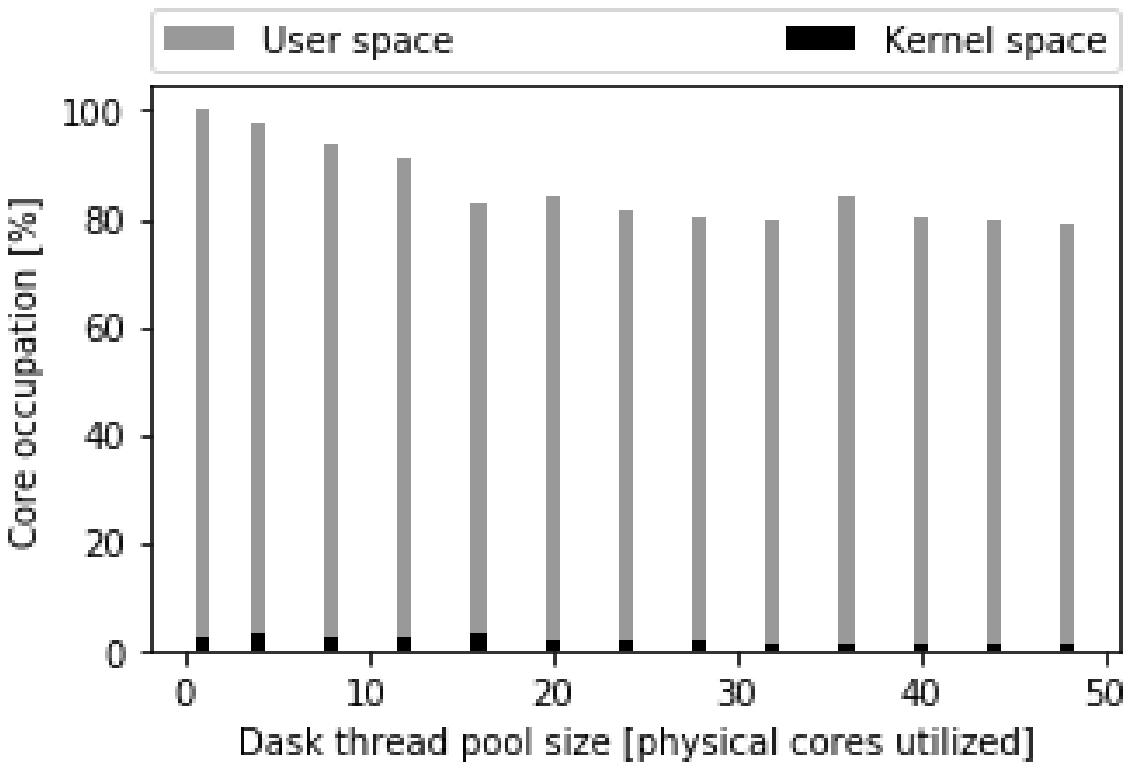}
    \end{minipage}
    \begin{minipage}{0.36\textwidth}
         \centering
         \includegraphics[width=0.99\textwidth]{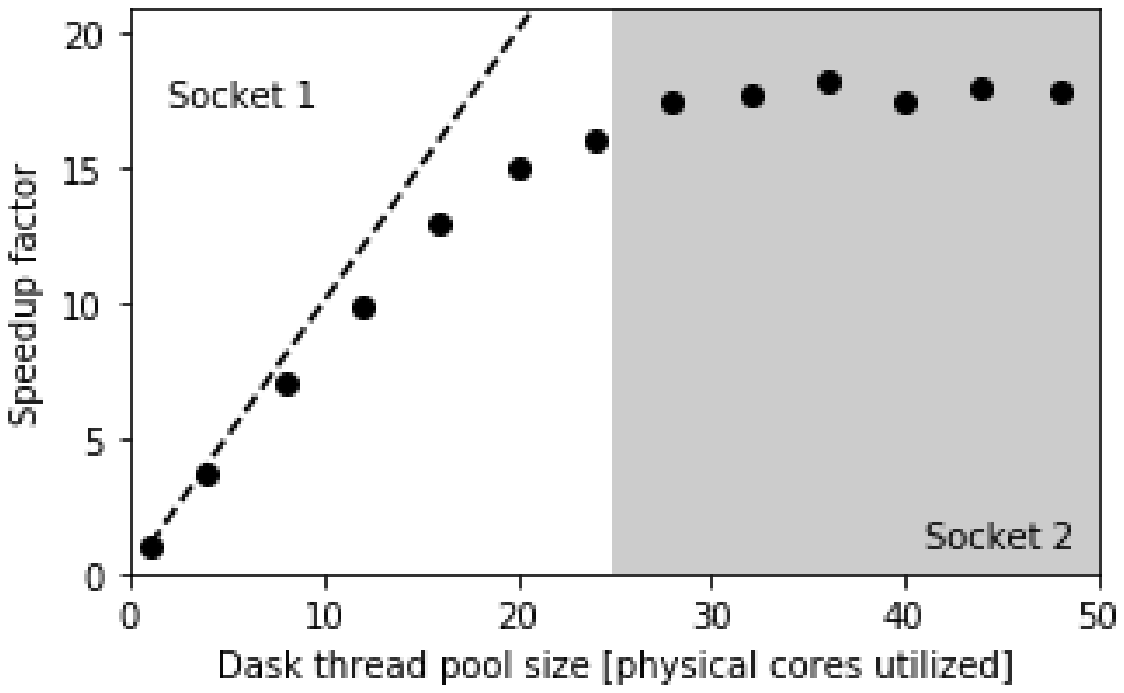}
    \end{minipage}
    \caption{Scaling performance of our flagging software after fine-tuning cache performance through the chunk size parameters in the user interface. Dashed line represents perfect linear scaling for reference. Overall core occupancy is good although scaling is ultimately limited by the memory intensity of the underlying algorithm.}
    \label{fig:scaling}
\end{figure}

We profiled our implementation with a dual-socket Intel Xeon 8160 system with 550~GiB of memory applying a typical multi-pass L-band flagging strategy to a sufficiently large ($>100~\text{GiB}$) coarsely channelized (208kHz) dataset captured at a dump rate of 1s. After tuning thread affinity and chunk size to optimize cache performance (through non-intrusive L3 cache profiling with the Linux kernel utility \texttt{perf}) we show that our flagger scales well to more than 20 physical cores (Figure~\ref{fig:scaling}) and flags data at over 400~GiB/hr.

\section{Conclusions}
We have built a highly configurable and scalable RFI flagger for the MeerKAT radio telescope, that is capable of processing coarsely channelized wideband data well above line rate with typical flagging strategies. Our flagging software has substantially cut down on the processing time needed for telescope commissioning, as well as recent continuum science observation processing through the \textsc{CARACal} \citep{P9-81_adassxxx} pipeline, as well as processing of the \textsc{Mightee} continuum survey \citep{delhaize2020mightee} through the \textsc{Oxkat} \citep{2020ascl.soft09003H} pipeline.

While Tricolour is currently restricted to operation on a single node, \dask offers the potential for horizontal scaling on a cluster, as described in \citet{D8-132_adassxxx}. As we support data ingest through the generic, \textit{defacto} standard, Measurement Set v2.0 interface, modification of the flagger to support instruments other than MeerKAT should not be difficult to accomplish.

\section*{Acknowledgements}
{\footnotesize
All figures in this paper were generated using Matplotlib - a 2D graphics package used for Python for application development, interactive scripting,and publication-quality image generation across user interfaces and operating systems.
The MeerKAT radio telescope is a facility operated by the South African Radio Astronomy Observatory. A business unit of the National Research Foundation of South Africa.
}





\bibliography{P10-244}


\end{document}